\documentstyle[multicol,epsf,aps]{revtex}
\begin{document}
\title{Comment on ``Dynamic Scaling in the Spatial Distribution 
of Persistent Sites''}
\author{E.~Ben-Naim$^1$ and P.~L.~Krapivsky$^2$}
\address{$^1$Theoretical Division and Center for Nonlinear Studies, 
Los Alamos National Laboratory, Los Alamos NM 87545} 
\address{$^2$Center for Polymer Studies and Department of Physics, 
Boston University, Boston MA, 02215}
\maketitle
\begin{multicols}{2}
Recently, Manoj and Ray \cite{mr} investigated the spatial
distribution of unvisited sites in the one-dimensional single-species
annihilation process $A+A\to 0$.  They claimed that this distribution
is characterized by a new length scale ${\cal L}(t)\sim t^z$, and that
the dynamical exponent $z$ depends upon the initial concentration. We
show numerically that this assertion is erroneous. Regardless of the
initial concentration, this spatial distribution is characterized by
the diffusive length scale ${\cal L}_D(t)\sim (Dt)^{1/2}$.

The spatial distribution of unvisited sites is described by $F_l(t)$,
the probability density that two consecutive unvisited sites are
separated by $l$ sites.  This quantity satisfies $P_0(t)=\sum_l
F_l(t)$ and $1=\sum_l (l+1) F_l(t)$, with $P_0(t)$ the fraction of
unvisited sites. The latter condition follows from length
conservation.

In the diffusive annihilation process $A+A\to 0$, the particle density
$n(t)$ decays algebraically according to $n(t)\simeq (8\pi
Dt)^{-1/2}$, where $D$ is the hopping rate. This behavior is
independent of the initial concentration. Furthermore, writing
$n(t)\sim 1/{\cal L}_D(t)$ suggests that the diffusive length is the
only asymptotically relevant length scale.  As shown in Fig.~1, the
following scaling form holds
\begin{equation}
\label{flt}
F_l(t)\sim t^{-1}{\cal F}\left({lt^{-1/2}}\right),
\end{equation}
indicating that $F_l(t)$ is characterized by ${\cal L}_D(t)$ alone.
The scaling form (\ref{flt}) is consistent with the normalization
$1=\sum_l (l+1) F_l(t)$.  We stress that the scaling function ${\cal
F}(x)$ is also independent of the initial conditions.

The known decay of the number of unvisited sites $P_0(t)\sim
t^{-\theta}$ ($\theta=3/8$ is the persistence exponent) together with
the requirement $P_0(t)=\sum_l F_l(t)$ can now be used to infer the
small $x$ divergence of ${\cal F}(x)$
\begin{equation} 
\label{fx}
{\cal F}(x)\sim x^{-2(1-\theta)}\qquad x\to 0. 
\end{equation}
In the complementary $x\to \infty$ limit, we observed an exponential
decay ${\cal F}(x)\sim \exp(-Ax)$, indicating independence of distant
unvisited sites.  Similar small argument divergence and exponential
tail underly a related quantity $P_n(t)$ \cite{blk}, the probability
that a site has been visited $n$ times.  In both cases, the
corresponding exponent follows directly from $\theta$. Furthermore,
Eqs.~(\ref{flt})-(\ref{fx}) extend to persistent sites in the
$q$-state Potts-Glauber model if $\theta$ is replaced by $\theta(q)$
\cite{dhp}.

In summary, the spatial distribution of unvisited sites is
characterized by the diffusive length scale. Although the initial
concentration affects the transient behavior, it is irrelevant
asymptotically (see Fig.~2). The scaling form suggested in
Ref.~\cite{mr} is correct only if the choices $z=1/2$,
$\omega=\theta$, and $\tau=2(1-\theta)=5/4$ are made.  Thus, no
additional independent exponents emerge from $F_l(t)$.

\begin{figure}
\centerline{\epsfxsize=7cm \epsfbox{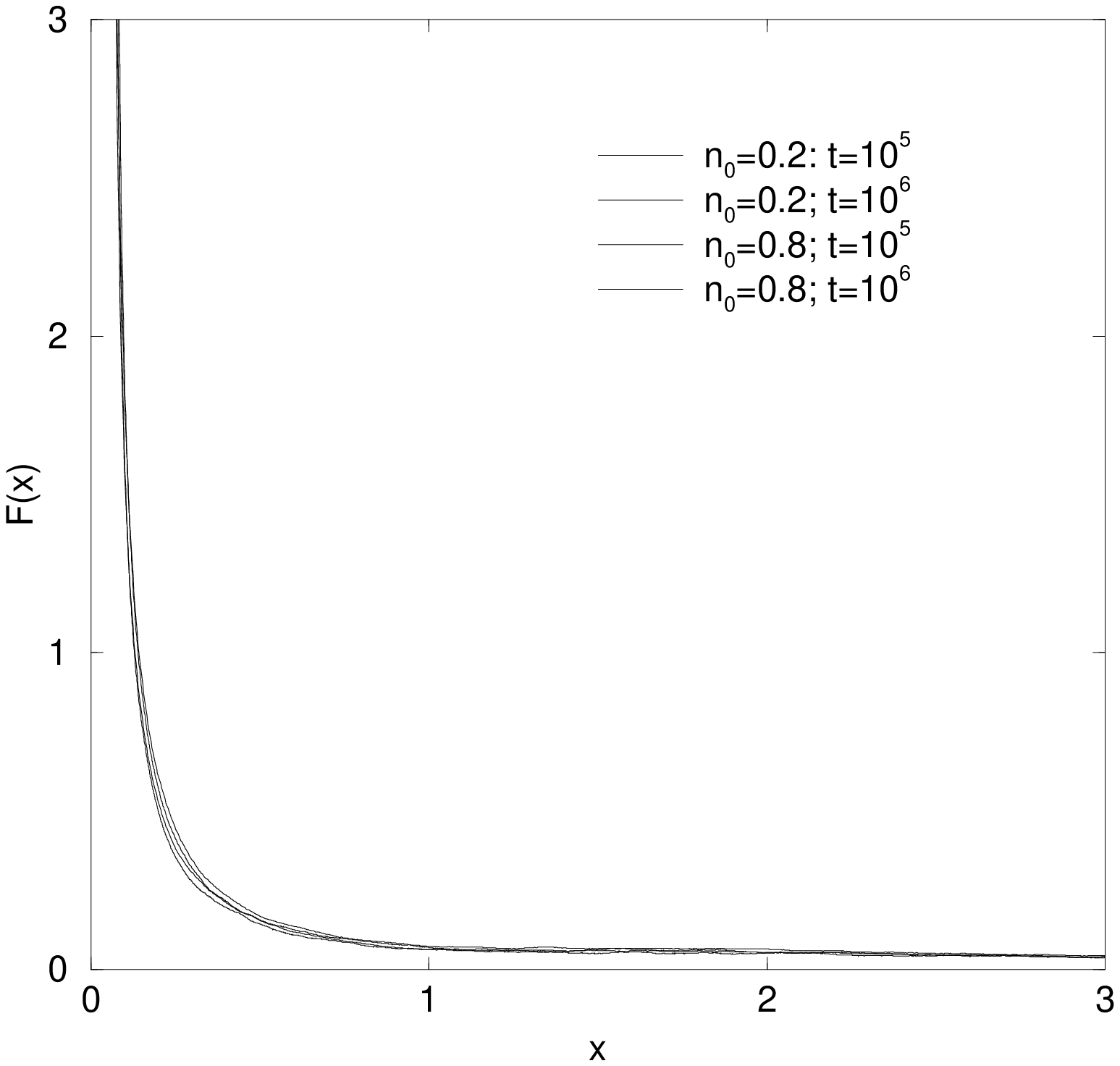}}
\noindent{\small {\bf Fig.~1} The scaling distribution ${\cal
F}(x)\equiv tF_l(t)$ versus $x=lt^{-1/2}$ for two different times
$t=10^5$, $10^6$, and two different initial densities $n_0=0.2$,
$0.8$.  Numerical simulations were performed in a system of size
$L=10^7$ and the results represent an average over 10 different
realizations.}
\end{figure}
\begin{figure}
\centerline{\epsfxsize=7cm \epsfbox{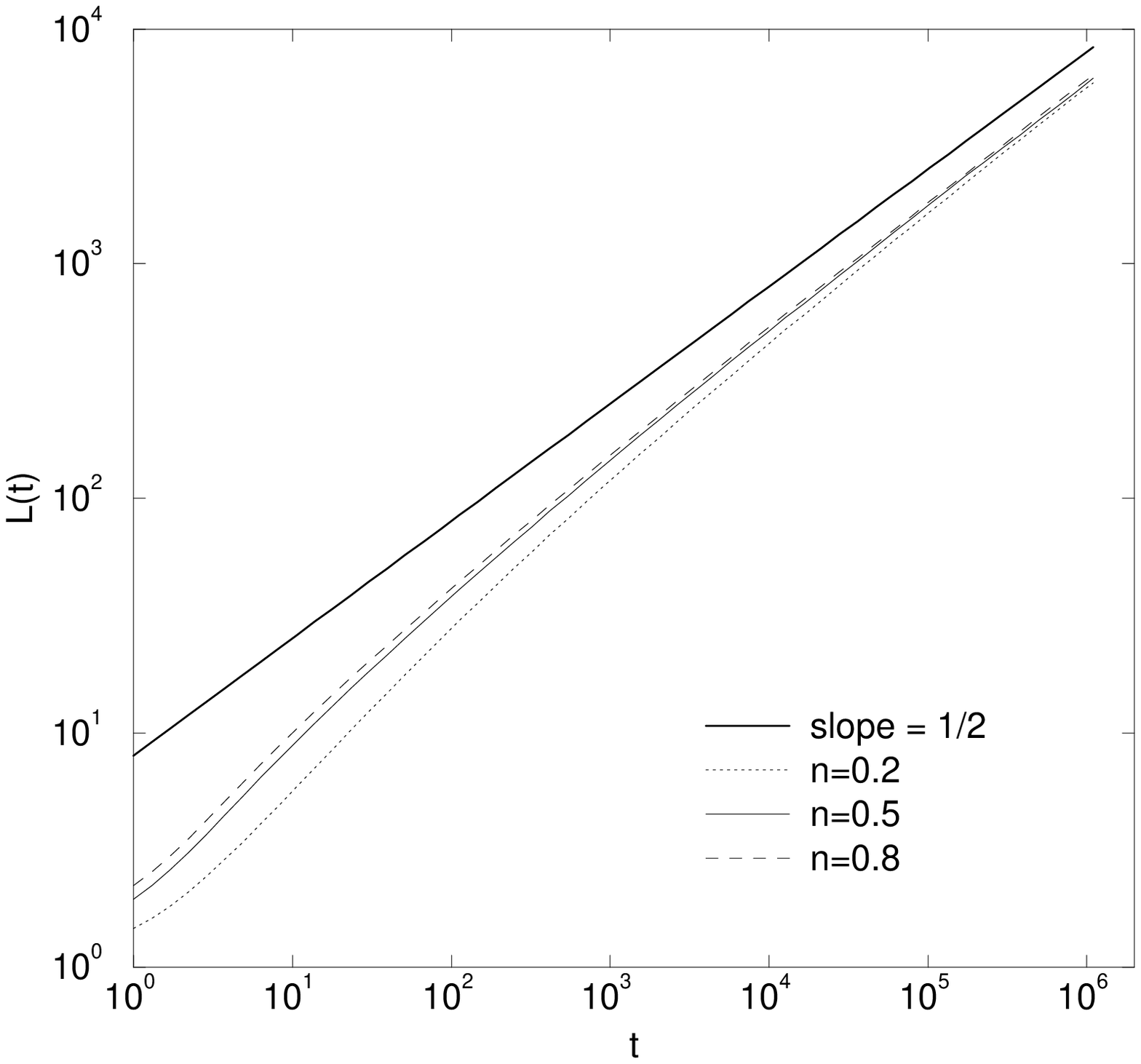}}
\noindent{\small {\bf Fig.~2} Time dependence of the typical domain
size \hbox{$L(t)=\sum_l l^2 F_l(t)/\sum_l l F_l(t)$}.}
\end{figure}

\end{multicols}

\end{document}